\documentclass[aps,prl,twocolumn,longbibliography,superscriptaddress]{revtex4-2}
\usepackage{bbm}
\usepackage{graphicx}
\usepackage{dcolumn}
\usepackage{bm}
\usepackage{subfigure}
\usepackage{amsmath}
\usepackage{feynmf}
\usepackage{hyperref}
\usepackage{amssymb}
\usepackage{attachfile}
\usepackage{multirow}
\usepackage{makecell}
\usepackage{ulem}
\newcount\supplementpage
\newcount\supplementpages
\newcommand{\appendSupplementPDF}[1]{%
  \clearpage
  \begingroup
  \hoffset=-1in
  \voffset=-1in
  \pdfximage{#1}%
  \supplementpages=\pdflastximagepages
  \supplementpage=1
  \loop
    \pdfximage width\paperwidth height\paperheight
      page\supplementpage{#1}%
    \shipout\vbox{\offinterlineskip\hbox{\pdfrefximage\pdflastximage}}%
    \advance\supplementpage by1
  \ifnum\supplementpage<\numexpr\supplementpages+1\relax
  \repeat
  \endgroup
}

\newcommand{\bk}{\boldsymbol k}

\newcommand{\bN}{\boldsymbol{N}}
\newcommand{\bs}{\boldsymbol{s}}
\newcommand{\ba}{\boldsymbol a}
\newcommand{\bb}{\boldsymbol b}
\newcommand{\bK}{\textbf K}
\newcommand{\bq}{\boldsymbol q}

\begin{document}

\title{Odd-Parity Altermagnetism Originating from Orbital Order}

\author{Zheng-Yang Zhuang}
\email{zhuangzhy3@gmail.com}
\affiliation{Guangdong Provincial Key Laboratory of Magnetoelectric Physics and Devices,
	State Key Laboratory of Optoelectronic Materials and Technologies,
	School of Physics, Sun Yat-sen University, Guangzhou 510275, China}
\affiliation{Department of Physics, Hong Kong University of Science and Technology, Clear Water Bay, Hong Kong, China}

\author{Di Zhu}
\affiliation{Guangdong Provincial Key Laboratory of Magnetoelectric Physics and Devices,
	State Key Laboratory of Optoelectronic Materials and Technologies,
	School of Physics, Sun Yat-sen University, Guangzhou 510275, China}

\author{Dongling Liu}
\affiliation{Guangdong Provincial Key Laboratory of Magnetoelectric Physics and Devices,
	State Key Laboratory of Optoelectronic Materials and Technologies,
	School of Physics, Sun Yat-sen University, Guangzhou 510275, China}

\author{Zhigang Wu}
\affiliation{Quantum Science Center of Guangdong-Hong Kong-Macao Greater Bay Area (Guangdong), Shenzhen 508045, China}

\author{Zhongbo Yan}
\email{yanzhb5@mail.sysu.edu.cn}
\affiliation{Guangdong Provincial Key Laboratory of Magnetoelectric Physics and Devices,
	State Key Laboratory of Optoelectronic Materials and Technologies,
	School of Physics, Sun Yat-sen University, Guangzhou 510275, China}

\date{\today}

\maketitle

\section*{Summary}
Odd-parity spin splitting plays a central role in spintronics and unconventional superconductivity, yet its microscopic realization in collinear magnetic systems remains elusive. We propose a general symmetry-based strategy for realizing odd-parity altermagnetism by stacking two noncentrosymmetric monolayers in an interlayer antiferromagnetic configuration. In this setting, odd-parity spin splitting originates from nonrelativistic complex orbital order rather than spin-orbit coupling.
By exploiting lattice symmetries, our framework enables the realization of both $p$- and $f$-wave altermagnets with nontrivial topological properties. Depending on the symmetry principles governing the bilayer stacking, the resulting altermagnet can host either quantum spin Hall or Chern insulator phases.
Our work expands the landscape of altermagnetic phases and opens new avenues for discovering fascinating physical phenomena.

\section*{KEYWORDS}

Altermagnetism, odd-parity spin splitting, quantum spin Hall insulators, layer Hall effects

\section*{INTRODUCTION}

Momentum-dependent spin splitting (MDSS) provides a fundamental link between the geometry of Bloch states and a diverse range of topological phases and spin transport phenomena. Based on the momentum-space symmetry of $E_{\bk,s}$, MDSS can be broadly classified into three classes~\cite{Hayami2019NRSS,Hayami2020NRSS,Yuan2021AM,Yuan2024PRL}: (i) even-parity, where $E_{\bk,s}=E_{-\bk,s}$, (ii) odd-parity, where $E_{\bk,s}=E_{-\bk,-s}$, and (iii) mixed-parity, where neither equality holds. Spin-orbit coupling (SOC) in time-reversal-invariant noncentrosymmetric materials is a prototypical origin of MDSS~\cite{Kane2005,Bernevig2006,Galitski2013}. Time-reversal symmetry then requires the resulting spin splitting to have odd parity. Over the past two decades, SOC-induced MDSS has stimulated extensive theoretical and experimental efforts~\cite{Galitski2013,Sinova2015,Manchon2015,Schaffer2016,Soumyanarayanan2016,Smidman2017}, including the discovery of various topological phases and exotic electromagnetic responses~\cite{Fu2008TSC,Sato2009TSC,Lutchyn2010,Oreg2010,Yu2010QAHE,Chang2013QAHE}.

Recently, a novel class of collinear antiferromagnets known as altermagnets~\cite{Mazin2023AM,Libor2022AMa} has drawn wide interest for hosting nonrelativistic spin-split bands without net magnetization~\cite{Libor2022AMa,Libor2022AMb,Ma2021AM,wu2007,Yuan2020AM,Yuan2021AM,Libor2020AM,Libor2022AMc,Osumi2024MnTe,Lee2024MnTe,Krempasky2024,Hajlaoui2024AM,Reimers2024,Ding2024CrSb,Yang2024CrSb,Zeng2024CrSb,Li2024CrSb,Lu2024AM7,Jiang2024KV2Se2O,Bai2024review,jungwirth2025,Liu2025fFIM,guo2025FIM,Feng2025FIM}
and for their diverse associated phenomena~\cite{Rafael2021AM,Ouassou2023AM,Bai2023AM,Fang2023NHE,Zhu2023TSC,Lu2024AM11,Zhang2024AM,Zhu2024dislocation,Ghorashi2024AM,Jin2024AM11,Attias2024AM9,Parshukov2024AM,Han2024AM,Brekke2023AM,Wei2024AM9,Jiang2024AM12,
	Antonenko2024AM,Hu2025NLME,Hu2025Cpair,Duan2025AFMAM,Zhang2025Cpair,Camerano2025,Lin2025AM4,Chen2025AM7}.
Within spin-space group theory~\cite{Libor2022AMa,Libor2022AMb,Liu2022AM,Liu2024AMPRX,Jiang2024SSG,Xiao2024SSG}, they exhibit even-parity MDSS~\cite{Libor2022AMa} (e.g., $d$-, $g$-, and $i$-wave), protected by an intrinsic spinless time-reversal symmetry $[\bar{C}_{2}\Vert\mathcal{T}]$ that imposes $E_{\bk,s}=E_{-\bk,s}$. Here, $\bar{C}_{2}$ is a $180^{\circ}$ spin rotation about an axis perpendicular to the ordered moments, combined with spin-space inversion~\cite{Libor2022AMa,Litvin1974,Litvin1977,Andreev1980}, and $\mathcal{T}$ denotes real-space time reversal.
In contrast, the realization of nonrelativistic odd-parity MDSS has received far less attention. So far, it has been discussed mainly in noncollinear coplanar magnets~\cite{Hellenes2023pwave,Brekke2024pwave,Yu2025Odd,Song2025Odd,yamada2025pwave}. Representative constructions of these states involve a lattice translation combined with either a spin rotation or time reversal. Because the local exchange field changes direction from site to site, no single global spin projection is generally conserved even without SOC. A more recent idea introduces odd-parity MDSS in bipartite collinear antiferromagnets~\cite{lin2025}, where sublattice currents generate complex hoppings akin to those in the Haldane model~\cite{Haldane1988}. This breaks $[\bar{C}_{2}\Vert\mathcal{T}]$ but preserves $[C_{2}\Vert\mathcal{P}]$~\cite{zeng2025}, where $\mathcal{P}$ is spatial inversion. Because $[C_{2}\Vert\mathcal{P}]$ enforces $E_{\bk,s}=E_{-\bk,-s}$, it guarantees odd-parity MDSS in a collinear setting; systems realizing this condition are referred to as {\it odd-parity altermagnets}~\cite{lin2025}. Since their MDSS closely mimics the SOC-induced MDSS, these systems are expected to host a rich variety of interesting phenomena. Yet, the symmetry principles underlying odd-parity altermagnets remain far from fully understood. Whether odd-parity altermagnetism can be protected by a different symmetry remains an open question, and establishing a general route to such phases is therefore of fundamental importance.

Because $[\bar{C}_{2}\Vert\mathcal{T}]$ acts as spinless time-reversal symmetry, breaking it requires orbital magnetization. In the previous construction, sublattice currents, which represent real-space circulating currents, generate orbital magnetization, thereby breaking $[\bar{C}_{2}\Vert\mathcal{T}]$ and allowing odd-parity altermagnetism~\cite{lin2025}. Complex orbital order provides a distinct route. It consists of an ordered pattern of nonzero local orbital magnetizations, thereby breaking real-space (spinless) time-reversal symmetry. Complex orbital order was theoretically predicted to be stabilized by the interplay of on-site Coulomb and Hund interactions over finite parameter regimes in $(111)$-oriented LaNiO$_3$ bilayers embedded in insulating oxide heterostructures~\cite{Yang2011,Ruegg2011}. We also provide a separate proof-of-principle interaction mechanism in Note S2 of the Supplemental information (SI). Here, we take complex orbital order as an input that breaks spinless time-reversal symmetry. Its microscopic stabilization depends on the lattice structure and interactions.

In this work, we establish a bilayer stacking framework for realizing odd-parity altermagnetism driven by complex orbital order. Guided by symmetry, different stacking symmetries within this framework can generate distinct classes of odd-parity altermagnets. In particular, besides the previously discussed $[C_{2}\Vert\mathcal{P}]$-protected class, we identify a distinct two-dimensional class protected by the effective time-reversal symmetry $[C_{2}\Vert\mathcal{M}_{z}\mathcal{T}]$, which is the central focus of this work. Here, $\mathcal{M}_{z}$ exchanges the layers, while $\mathcal{T}$ reverses the local orbital moments; the combined symmetry enforces $E_{\bk,s}=E_{-\bk,-s}$. Compared with SOC-driven and coplanar mechanisms, the present construction is simultaneously nonrelativistic and collinear, with the spin component along the Néel vector conserved in the SOC-free limit. We construct explicit tight-binding models for $p$- and $f$-wave altermagnets and discover that they can host quantum spin Hall (QSH) insulator phases. This differs fundamentally from previous studies on systems with $[C_{2}\Vert\mathcal{P}]$, where one obtains Chern insulator phases~\cite{lin2025}.

\section*{RESULTS}

\subsection*{Symmetry analysis and general framework}
Generating odd-parity altermagnetism requires
three key conditions. First, the system must exhibit a compensated collinear magnetic order~\cite{Libor2022AMa}. 
Second, the bands must satisfy $E_{\boldsymbol{k},s}=E_{-\boldsymbol{k},-s}$. Earlier work showed that $[C_{2}\Vert\mathcal{P}]$ guarantees this condition; here we find that an effective time-reversal symmetry can play the same role. Third,
the bands must satisfy the non-even-parity condition $E_{\boldsymbol{k},s}\neq E_{-\boldsymbol{k},s}$ for generic momenta $\boldsymbol{k}$. 
In collinear antiferromagnets, this condition implies the breaking of inversion symmetry within each spin sector, 
as well as the breaking of  $[\bar{C}_{2}\Vert\mathcal{T}]$~\cite{Libor2022AMa}.
By contrast, in even-parity altermagnets, spin splitting can coexist with preserved $[\bar{C}_{2}\Vert\mathcal{T}]$ and inversion symmetries.

To fulfill the three conditions simultaneously, we propose the bilayer stacking strategy illustrated in Figure 1. We begin with two ferromagnetic monolayers stacked in an interlayer antiferromagnetic configuration. To satisfy the third condition, we consider noncentrosymmetric monolayers, since inversion symmetry would otherwise enforce $E_{\bk,s}=E_{-\bk,s}$. In addition, complex orbital order---or another mechanism that generates orbital magnetization $\vec{M}^{\rm orb}$---must be present to break $[\bar{C}_{2}\Vert\mathcal{T}]$.
We then flip the top layer by a twofold rotation $C_{2i}$ about an in-plane mirror line $i$ of the underlying noncentrosymmetric monolayer, so that the lattice remains invariant.
For a two-dimensional system, orbital magnetization is generally oriented perpendicular to the plane and is characterized by $\mathcal{L}_{z}$. This operation reverses both $\vec{M}^{\rm orb}=M_{z}^{\rm orb}\hat{z}$ and $\mathcal{L}_{z}$.
The resulting top layer is a time-reversed copy of the bottom layer, and the bilayer preserves $[C_{2}\Vert\mathcal{M}_{z}\mathcal{T}]$.

This symmetry enforces $E_{\bk,s}=E_{-\bk,-s}$. Together with the non-even-parity condition above, it yields nonzero odd-parity spin splitting. Because the Fermi-surface geometry is dictated by lattice symmetry, lattice design provides a route to engineer different forms of odd-parity altermagnetism.
For example, $p$- and $f$-wave altermagnets arise when the monolayer belongs to the $[E\Vert C_{1z}]$ and $[E\Vert C_{3z}]$ symmetry classes, respectively.

\subsection*{Models for $p$- and $f$-wave altermagnets}

We now implement the proposed framework in explicit lattice models. Guided by the symmetry requirements above, the minimal setup for a $[C_{2}\Vert\mathcal{M}_{z}\mathcal{T}]$-protected odd-parity altermagnet consists of two layers, each containing two sublattices with complex orbital order. Within this lattice-based construction, the minimal model therefore has eight bands, and its Hamiltonian reads
\begin{eqnarray}
	\mathcal{H}(\bk)=\left[\begin{array}{cc}
		\mathcal{H}_{t}(\bk)+J\bN\cdot\bs  & T_{\perp}(\bk)\\
		T_{\perp}^{\dagger}(\bk)& \mathcal{H}_{b}(\bk)-J\bN\cdot\bs \\
	\end{array}\right],\label{eq: model}
\end{eqnarray}
where $\bs=(s_{x},s_{y},s_{z})$ denotes the vector of Pauli matrices acting on spin space.
For notational simplicity, we omit all identity matrices throughout the paper.
The intralayer Hamiltonians of the top and bottom layers satisfy $\mathcal{H}_{b}(\bk)=\mathcal{H}_{t}^{*}(-\bk)$, as required by $[C_{2}\Vert\mathcal{M}_{z}\mathcal{T}]$.
The parameter $J$ is the exchange field arising from the intralayer ferromagnetic and interlayer antiferromagnetic order, $\bN$ is the Néel-vector direction, and $T_{\perp}(\bk)$ is the interlayer hopping matrix. Throughout, we choose the spin-quantization axis $z$ along the Néel vector. For the AA-stacked bilayer considered here, we retain the leading scalar, orbital-independent interlayer hopping $T_\perp=t_\perp$ for simplicity. Introducing more general symmetry-allowed orbital- and sublattice-dependent hoppings does not modify the symmetry or the resulting odd-parity altermagnetism.

To make the analysis transparent, we consider the simplest case where each sublattice in a monolayer hosts a pure atomic-orbital state. This choice is representative rather than unique.
The generic Hamiltonian for the top layer then becomes
\begin{eqnarray}
	\mathcal{H}_{t}(\bk)&=&\sum_{j}t_{j}\left[{\rm Re}(\mathcal{F}_{j,\bk})\sigma_{x}-{\rm Im}(\mathcal{F}_{j,\bk})\sigma_{y}\right]\nonumber\\
	&&+\sum_{j}\left(t_{d}\sigma_{z}+t_{0}\right)\cos(\bk\cdot\bb_{j})+\delta\sigma_{z},\label{eq: kinetic}
\end{eqnarray}
where $\mathcal{F}_{j,\bk}\equiv e^{i\bk\cdot\ba_{j}}e^{im\varphi_{j}}$ encodes the direction-dependent phase acquired by hopping between complex orbitals whose magnetic quantum numbers differ by $m$~\cite{supplemental}, and Pauli matrices $\sigma_{x,y,z}$ act on sublattices. The first row describes inter-sublattice hopping along the bond vectors $\ba_{j}$, whereas the second row represents intra-sublattice hopping along $\bb_{j}$ parameterized by $\{t_{0},t_{d}\}$ and the on-site potential difference $\delta$ between the two sublattices.
We emphasize that, at the tight-binding level, where spatial-symmetry breaking is encoded by the lattice, the present eight-band model is minimal by band counting for realizing $[C_{2}\Vert\mathcal{M}_{z}\mathcal{T}]$-protected odd-parity altermagnetism: the noncentrosymmetric monolayer and bilayer structure require both sublattice and layer degrees of freedom, in addition to spin.

For $p$-wave altermagnets, a simple realization is provided by a noncentrosymmetric square lattice with sublattice dimerization.
Here we assume that the dimerization is along the $x$ direction and that the two sublattices in the top (bottom) layer host the $p_{+}$ ($p_{-}$) and $s$ orbitals, respectively, where $p_{\pm}\equiv p_{x}\pm ip_{y}$ [Figure 2(a)], such that $|m|=1$. The bond vectors and their associated phases are listed in Table 1. The resulting Hamiltonian can be expressed in terms of Pauli matrices as
\begin{eqnarray}
	\mathcal{H}_{p}(\bk)&=&h_{1,\bk}^{p}(t_{0}+t_{d}\sigma_{z})+h_{2,\bk}^{p}\sigma_{x}+h_{3,\bk}^{p}\sigma_{y}+\delta\sigma_{z}\nonumber\\
	&&+(h_{4,\bk}^{p}\sigma_{y}+h_{5,\bk}^{p}\sigma_{x}+J\bN\cdot\bs)\rho_{z}+t_{\perp}\rho_{x},\label{eq: p-wave}
\end{eqnarray}
where the Pauli matrices $\rho_{x,y,z}$ act on the two layers.
The explicit forms of the functions $h_{i,\bk}^{p}$ are summarized in Table 2. 
In this configuration, the combined mirror operation $\mathcal{M}_{y}\mathcal{M}_{z}$ relates the two layers and enforces spin degeneracies along $\mathcal{M}_{y}$-invariant lines in the Brillouin zone [Figures 2(c) and 2(d)].
By contrast, if the dimerization is oriented along $y$, the relevant symmetry becomes $\mathcal{M}_{x}\mathcal{M}_{z}$, resulting in a $\pi/2$ rotation of the spin-splitting pattern [see Note S3 in SI]. Hence, the orientation of the monolayer dimerization provides an efficient knob to control the $p$-wave spin splitting.

\subsection*{Altermagnetic quantum spin Hall insulators}

In our framework, both the $p$- and $f$-wave models can realize spin-Chern phases that become QSH insulators when the global indirect gap is positive. The intuition is most transparent in the limit of vanishing interlayer coupling: the two layers decouple into independent ferromagnetic monolayers that each break time-reversal symmetry. When tuned into the topological regime, each monolayer becomes a Chern insulator. Because the two layers are related by an effective time-reversal symmetry, their Chern numbers cancel, yielding a zero total charge Chern number but a finite spin Chern number~\cite{Sheng2006}. This mechanism differs from QSH phases discussed in even-parity altermagnets, whose spectra are not generally constrained by $E_{\bk,s}=E_{-\bk,-s}$.

To make the discussion concrete, we focus on the $f$-wave model (the $p$-wave case is presented in Note S3 in SI). Setting $t_{\perp}=0$ renders the layer index $\rho$ (the eigenvalues of $\rho_{z}$) a good quantum number. The Hamiltonian then decomposes into four $2\times2$ sectors $\mathcal{H}_{f}^{s,\rho}$ labeled by spin $s$ and layer index $\rho$. Without loss of generality, we assume both $\delta$ and $t_{d}$ are positive. In the large-$\delta$ limit, all sectors are topologically trivial. Adiabatically decreasing $\delta$ triggers gap closings and reopenings at the $\bK$ valley for the $\rho=1$ sectors and at the $\bK^{\prime}$ valley for the $\rho=-1$ sectors, signaling a topological phase transition.
Expanding near the valleys gives
\begin{eqnarray}
	\mathcal{H}_{f,\tau}^{s, \rho}(\bq)=m^{s,\rho}+(\delta-\frac{3t_{d}}{2})\sigma_{z}+\frac{3t}{2}(\tau q_{x}\sigma_{x}+q_{y}\sigma_{y}),
	\label{eq: f-wave-valley}
\end{eqnarray}
with $m^{s,\rho}=s\rho J-\frac{3t_{0}}{2}$, and $\tau=\pm1$ for $\bK$, $\bK^{\prime}$. The vector $\bq=(q_{x},q_{y})$ is the momentum relative to the valleys. Each low-energy Hamiltonian describes a massive Dirac cone carrying a Chern number $\mathcal{C}^{s}_{\rho,\tau=\rho}={\rm sgn}[\tau(\tfrac{3t_{d}}{2}-\delta)]/2$. As $\delta$ decreases across the critical value $\delta=3t_{d}/2$, 
the Chern number $\mathcal{C}^{s}_{\rho}$ for the sector $\mathcal{H}_{f}^{s,\rho}$
jumps from zero to $\mathcal{C}^{s}_{\rho}={\rm sgn}(\rho t_{d})$, rendering $\mathcal{H}_{f}^{s,\rho}$ a Chern insulator. Because opposite layers carry opposite Chern numbers, the upper direct gap $G_+$ separates their conduction bands, whereas the lower direct gap $G_-$ separates their valence bands [Figure 3(a)]; each gap therefore defines a topologically nontrivial phase.
For $t_{0}=0$, this occurs when the exchange field $J$ exceeds the bandwidth $W$ [see the $t_{\perp}=0$ line in Figure 3(b)].
For the full system including both spin sectors, the total Chern number always vanishes due to the effective time-reversal symmetry $[C_{2}\Vert\mathcal{M}_{z}\mathcal{T}]$ that connects opposite spins. Nevertheless, the spin Chern number~\cite{Sheng2006} remains finite within the gaps $G_{\pm}$:
\begin{equation}
	\mathcal{C}^{S}_{G_{\pm}}=\frac{\mathcal{C}^{\uparrow}_{\rho=\pm1}-\mathcal{C}^{\downarrow}_{\rho=\mp1}}{2}=\pm{\rm sgn}(t_{d}).
\end{equation}
The system therefore hosts a pair of helical edge states traversing each relevant direct gap. This topology persists at finite $t_{\perp}$ as long as the relevant direct gap remains open, as shown in Figure 3(b).
The role of $t_{0}$ warrants clarification. Because it enters the Hamiltonian through an identity matrix, it shifts all bands equally at fixed $\bk$ without changing their eigenvectors, Berry curvature, direct gaps, or the Chern number of a separated band subspace. Nevertheless, it can modify the indirect gap and render the system a topological metal or indirect semimetal without changing its band topology.

The protection mechanism differs from that of a conventional QSH insulator. A conventional QSH insulator is protected by physical time reversal $\mathcal{T}=is_y\mathcal{K}$, with $\mathcal{T}^2=-1$, where $\mathcal{K}$ denotes complex conjugation. Although magnetism breaks $\mathcal{T}$ here, the bilayer possesses the global antiunitary symmetry
$\widetilde{\mathcal{T}}=C_{2z}^{(s)}\mathcal{M}_{z}\mathcal{T}
=i\rho_xs_y\mathcal{K}$,
where $C_{2z}^{(s)}$ is a $\pi$ spin rotation about the $z$ axis. Because $\widetilde{\mathcal{T}}^2=-1$, it provides class-AII protection when a global insulating gap is open (see Note S7 in SI). Within the projected $\{s,p_x,p_y\}$ orbital subspace, the atomic SOC $\lambda_{\rm at}\mathcal{L}_{z}s_z$ preserves $\widetilde{\mathcal{T}}$ and hence the $\mathbb Z_2$ topology. Rashba SOC, by contrast, breaks $\widetilde{\mathcal{T}}$ and removes the global class-AII protection. However, it does not destory the QSH phase either, due to an emergent antiunitary symmetry
that protects the edge-state spectra crossing at the time-reversal invariant boundary momentum $k_{x}=0$ (see Note S7 in SI).

\subsection*{Layer Hall effect}
In addition to the spin Hall response driven by an in-plane electric field, an out-of-plane electric field provides a complementary electrical probe of the layer-exchanging symmetry protecting the odd-parity phase. Such a field introduces the layer-dependent potential $(eE_zd/2)\rho_z$ and breaks $[C_{2}\Vert\mathcal{M}_{z}\mathcal{T}]$, where $d$ is the interlayer distance. As a result, an electrically switchable anomalous Hall response becomes allowed: at zero field, $[C_{2}\Vert\mathcal{M}_{z}\mathcal{T}]$ enforces a vanishing total charge Hall conductance. In the weak-interlayer-coupling regime, this cancellation can be understood through a spin-layer-locking picture in which the opposite Hall contributions are predominantly localized in opposite layers. When the system is metallic, this layer potential shifts the occupations of the two sectors unequally, removes the cancellation of their Hall contributions, and produces a net charge Hall conductance satisfying $\sigma_{xy}(E_{z})=-\sigma_{xy}(-E_{z})$ [Figure 4(a)]. Owing to its layered origin, this effect is also called the layer Hall effect~\cite{Gao2021}. By contrast, in a conventional QSH system, physical time-reversal symmetry forbids a linear charge Hall response.

\subsection*{Effects of orbital mixing}
In realistic systems with complex orbital order, the electronic states at a given lattice site can be linear combinations of multiple atomic orbitals. Because the inter-orbital hoppings are essential for generating odd-parity MDSS and nontrivial topology, we examine how orbital mixing affects the altermagnetic phase.
When orbital mixing is present, all possible hoppings between occupied orbitals of different sublattices must in principle be considered.
Nevertheless, lattice symmetry selects which hopping channels can contribute to the odd-parity terms.
On a square lattice, for instance, the hopping phase takes the form $e^{im(j-1)\frac{\pi}{2}}$, which becomes purely real when $m=2n$ with $n \in \mathbb{Z}$ (since $e^{in\pi}=\pm1$). Thus, only hoppings with odd angular momentum difference ($m\in2\mathbb{Z}+1$) contribute to the odd-parity MDSS. A similar argument applies to systems with $C_{3}$ symmetry, where the phase becomes trivial for $m\in 3\mathbb{Z}$, and only hoppings with $m\notin 3\mathbb{Z}$ are relevant.

To illustrate this more concretely, we revisit the $f$-wave altermagnet shown in Figure 2(e). Suppose the purple sublattices in the top (bottom) layer now host a mixture of $s$ and $p_{-}$ ($p_{+}$) orbital states. This introduces additional hoppings between $p_{-}$ and $p_{+}$ orbitals, whose sole effect in the present parametrization is to renormalize the prefactors of $h_{4,\bk}^{f}$ and $h_{5,\bk}^{f}$ by a parameter $\eta$ that quantifies the degree of orbital mixing [see Note S5 in SI].
As $\eta$ decreases from $\eta=1$, the spin splitting diminishes and eventually vanishes at the critical point $\eta_{c}=0$, where an emergent $[C_{2}\Vert\mathcal{M}_{z}]$ symmetry completely suppresses the $f$-wave spin splitting~\cite{Zeng2024AM1,Zeng2024AM2}. For $\eta<\eta_{c}$, the spin splitting reappears with a reversed pattern, representing a time-reversed partner of the $\eta>0$ regime. This reversal is accompanied by a topological transition. As a direct manifestation, the spin Hall conductance~\cite{Sinova2015} changes sign across the transition, as shown in Figure 4(b).

\section*{DISCUSSION}

Guided by symmetry, we demonstrate that odd-parity altermagnetism can emerge from the interplay of complex orbital order and collinear antiferromagnetism. This mechanism yields QSH insulators when the global indirect gap is positive, as well as electrically switchable anomalous Hall responses in the metallic regime.

A complementary stacking procedure generates odd-parity altermagnets protected by $[C_{2}\Vert\mathcal{P}]$, which can realize Chern insulators with even Chern numbers already at small exchange fields (see Note S6 in SI). For both the $p$- and $f$-wave realizations of the $[C_{2}\Vert\mathcal{M}_{z}\mathcal{T}]$ construction, the $C_{2i}$ layer flip is performed about an in-plane mirror line of the underlying noncentrosymmetric monolayer, ensuring that its lattice remains invariant. The $\mathcal{L}_z$-type complex orbital order breaks the corresponding unitary mirror $\mathcal{M}_v$ and $\mathcal{T}$ separately while preserving $\mathcal{M}_v\mathcal{T}$. Together with the $C_{1z}$ and $C_{3z}$ rotational subgroups, this gives the constituent monolayers the magnetic point groups $m^{\prime}$ and $3m^{\prime}$, respectively. By contrast, the $[C_{2}\Vert\mathcal{P}]$ construction uses a $C_{2z}$ rotation and does not rely on the same in-plane layer flip. These magnetic point groups characterize the constituent monolayers, whereas the odd-parity spectral relation of the bilayer is enforced by either $[C_{2}\Vert\mathcal{M}_{z}\mathcal{T}]$ or $[C_{2}\Vert\mathcal{P}]$.

The layered structure of the altermagnets in our framework suggests that magnetic van der Waals materials~\cite{Amilcar2021CrCl3,Zhang2019MnBi2Te4,Li2019MnBi2Te4,Aapro2021MnSe,Zhang2019CrBr3,Lee2021CrSBr}, especially magnetic Janus monolayers~\cite{Liu2024AM}, provide a broad family of possible building blocks. Interaction-driven complex orbital order has been predicted in (111)-oriented LaNiO$_3$ bilayers embedded in insulating oxide heterostructures~\cite{Yang2011,Ruegg2011}. This suggests that such an oxide bilayer could serve as a monolayer building block in our stacking framework.
More broadly, our framework applies to monolayers with other nonrelativistic sources of orbital magnetization, such as chiral phonons~\cite{yao2025} or loop-current order~\cite{saati2025}.
Odd-parity spin splitting, the defining characteristic of odd-parity altermagnets, can be detected directly using spin-resolved ARPES~\cite{Lee2024MnTe,Osumi2024MnTe,Krempasky2024,Zhu2024}.

Protection by the layer-exchanging symmetry further distinguishes this phase from conventional QSH insulators. Time-reversal-even perturbations that break layer exchange---including layer-asymmetric strain, a piezoelectric layer potential, and asymmetric contacts---can shift the two sectors differently and produce layer-selective transport.

Beyond their fundamental interest, odd-parity altermagnets offer a potential platform for unconventional superconductivity. 
Although magnetism inherently breaks time-reversal symmetry, $E_{\bk,s}=E_{-\bk,-s}$ in odd-parity altermagnets provides a favorable kinematic condition for zero-center-of-mass pairing of opposite-spin states. This kinematic compatibility might lead to the coexistence of superconductivity and magnetism. Notably, recent studies on noncollinear $p$-wave magnets have shown that the interplay between $p$-wave spin splitting and superconductivity can lead to exotic phenomena~\cite{sun2025,Nagae2025pwave,Soori2025pwave,Sukhachov2025pwave,Ezawa2024pwave,Wang2025pwave,zeng2025pwave,salehi2025pwave,fukaya2025pwave}.
Given that odd-parity altermagnets host similar Fermi-surface structures but within a simpler collinear magnetic setting, their connection to superconductivity is an interesting direction to explore.

In summary, we have developed a general symmetry-based framework for realizing odd-parity altermagnetism. By identifying $[C_{2}\Vert\mathcal{M}_{z}\mathcal{T}]$ as a distinct symmetry that protects odd-parity altermagnetism, our work significantly broadens the landscape of altermagnetic phases and opens new avenues for research in spintronics.

\subsection*{Note added}
During the preparation of this paper, we became aware of other independent works~\cite{li2025floquet,huang2025oddparityAM,zhu2025floquet,liu2025floquet} that discuss the generation of odd-parity spin splitting in 2D and 3D collinear antiferromagnets. Their approach relies on Floquet engineering to generate orbital magnetizations. In all of these settings, odd-parity spin splitting is enforced by $[C_{2}\Vert\mathcal{P}]$, rather than by the $[C_{2}\Vert\mathcal{M}_{z}\mathcal{T}]$ symmetry emphasized here.

\section*{METHODS}
Details regarding the methods can be found in the Supplemental information.



\section*{RESOURCE AVAILABILITY}


All data needed to evaluate the conclusions in the paper are present in the paper
and/or the supplemental information. Additional data related to this paper
are available from the authors upon reasonable request.

\subsection*{Lead contact}


Requests for further information and resources should be directed to and will be fulfilled by the lead contact, Zhongbo Yan (yanzhb5@mail.sysu.edu.cn).

\subsection*{Materials availability}


This study did not generate any new unique materials.

\subsection*{Data and code availability}


\begin{itemize}
    \item All data necessary to evaluate the conclusions of this paper are included in the main text and/or the Supplemental Information.
    \item Any additional information required to reanalyze the data reported in this paper is available from the lead contact upon request.    
\end{itemize}

\section*{ACKNOWLEDGMENTS}


Z.-Y.Z. would like to thank Song-Bo Zhang, Luigi Camerano, and Jin-Xin Hu for helpful discussions, and the organizers of the ``Workshop on altermagnets'' for providing valuable opportunities for academic exchange.
This work was funded by the Fundamental and Interdisciplinary Disciplines Breakthrough
Plan of the Ministry of Education of China (Grant No.
JYB2025XDXM403), National Natural Science Foundation of China via grant No. 12174455, No. 12474264, Guangdong Basic and Applied Basic Research Foundation via grant No. 2023B1515040023, Guangdong Provincial Quantum Science Strategic Initiative via grant No. GDZX2404007, and National Key R\&D Program of China via grant No. 2022YFA1404103.

\section*{AUTHOR CONTRIBUTIONS}

Z.-Y.Z. and Z.Y. conceived the project; Z.-Y.Z. performed the calculations, and drew the figures; Z.-Y.Z., Z.W., and Z.Y. wrote the manuscript with input from all authors; and Z.Wu, and Z.Yan were responsible for funding acquisition.

\section*{DECLARATION OF INTERESTS}

The authors declare no competing interests.

\section*{SUPPLEMENTAL INFORMATION INDEX}
Document S1. Notes S1-S7, Figures S1-S6, and Table S1.

\section*{MAIN FIGURE TITLES AND LEGENDS}

\noindent\includegraphics[width=0.85\linewidth]{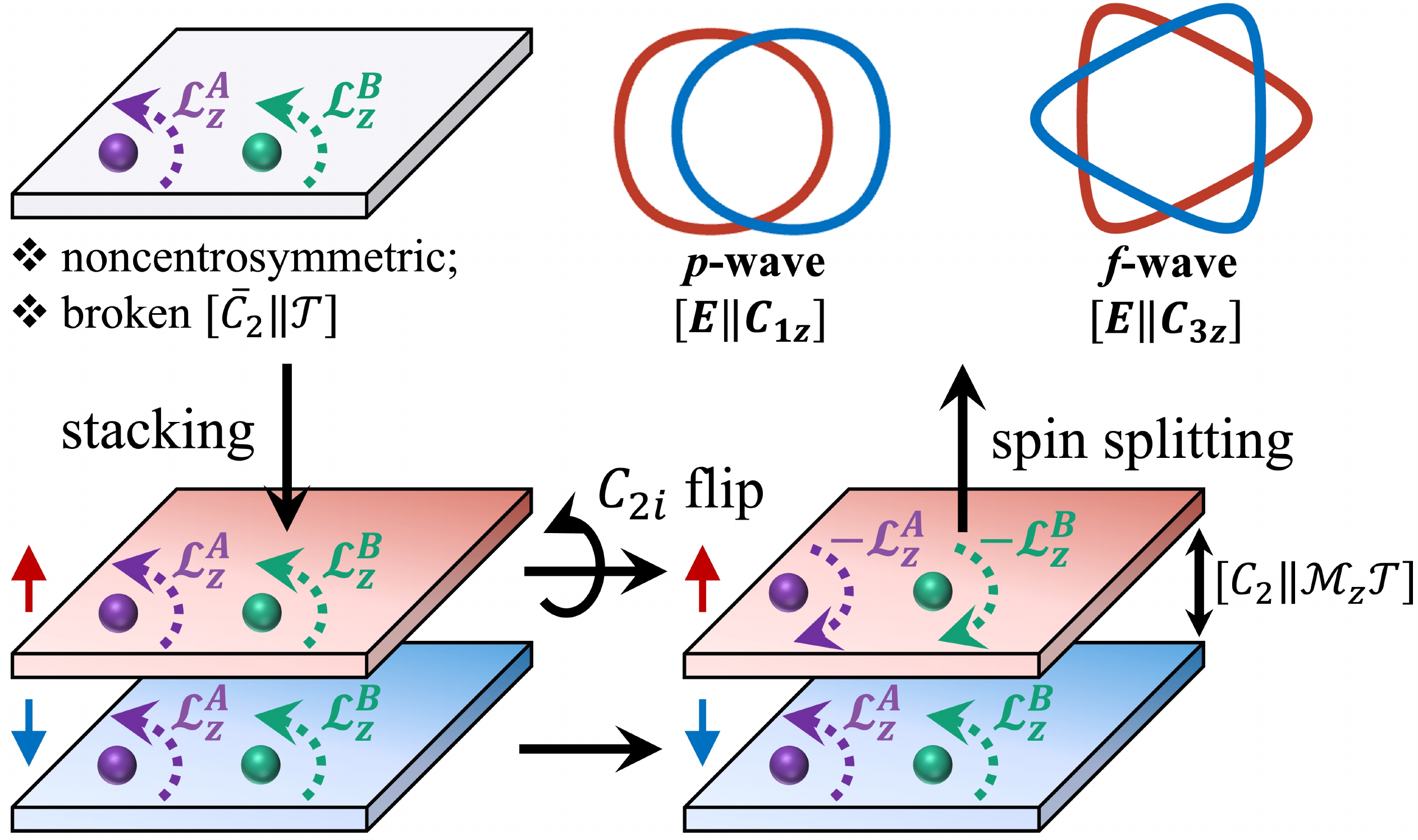}
\subsection*{Figure 1. Illustration of the framework}
We begin with a noncentrosymmetric monolayer containing at least two electronic states with different orbital angular momenta $\mathcal{L}_{z}^{A}$ and $\mathcal{L}_{z}^{B}$. Two such monolayers are stacked to form a bilayer exhibiting antiferromagnetic order. The top layer is then flipped by an in-plane $C_{2i}$ rotation, which reverses $\mathcal{L}_{z}^{A/B}$ while keeping the lattice unchanged. This procedure preserves $[C_{2}\Vert\mathcal{M}_{z}\mathcal{T}]$ but breaks both inversion symmetry and the real-space mirror symmetry $\mathcal{M}_{z}$, thereby giving rise to odd-parity spin splitting, as shown in the final stage.

\noindent\includegraphics[width=0.85\linewidth]{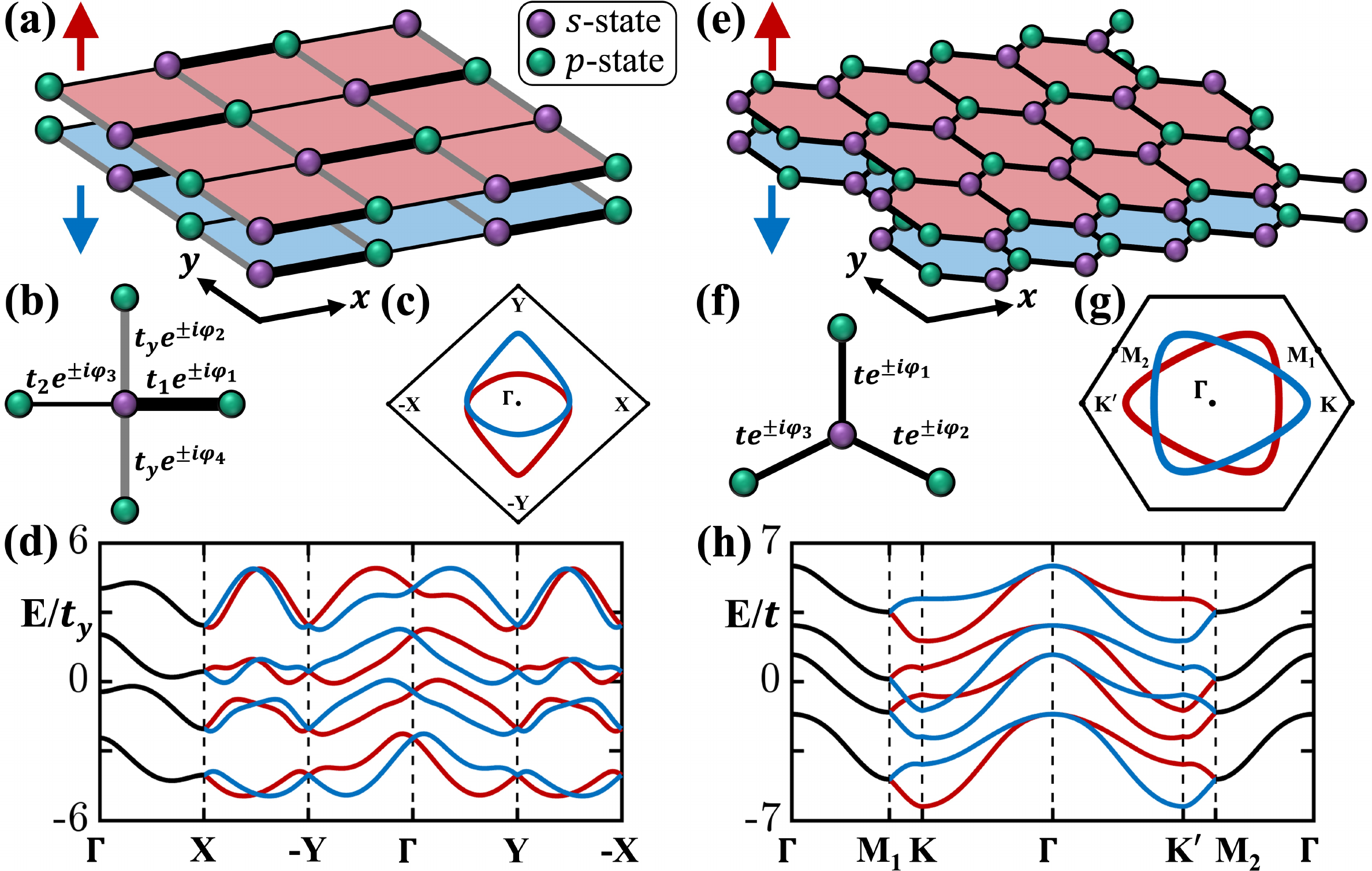}
\subsection*{Figure 2. Schematic of the lattice models for $p$- and $f$-wave altermagnets}
Panels (a) and (e) show the noncentrosymmetric square and hexagonal lattices used for $p$- and $f$-wave altermagnets, respectively. (b) and (f) illustrate the complex inter-sublattice hoppings. The sign $\pm$ in the exponential refers to the top/bottom layer. The resulting $p$- and $f$-wave patterns are displayed in (c-d) and (g-h), respectively. (c) [(g)] Spin-resolved Fermi surfaces at $E=-3.7t_{y}$ [$E=-3.5t$]. Common parameters for (c-d): $\{t_{1},t_{2},t_{y},t_{0},t_{d},t_{\perp},\delta,J\}=\{1.2,0.6,1,0.4,0.4,1,0,2\}$; for (g-h):
$\{t,t_{0},t_{d},t_{\perp},\delta,J\}=\{1,0.7,0.5,1,0,2\}$.

\noindent\includegraphics[width=0.85\linewidth]{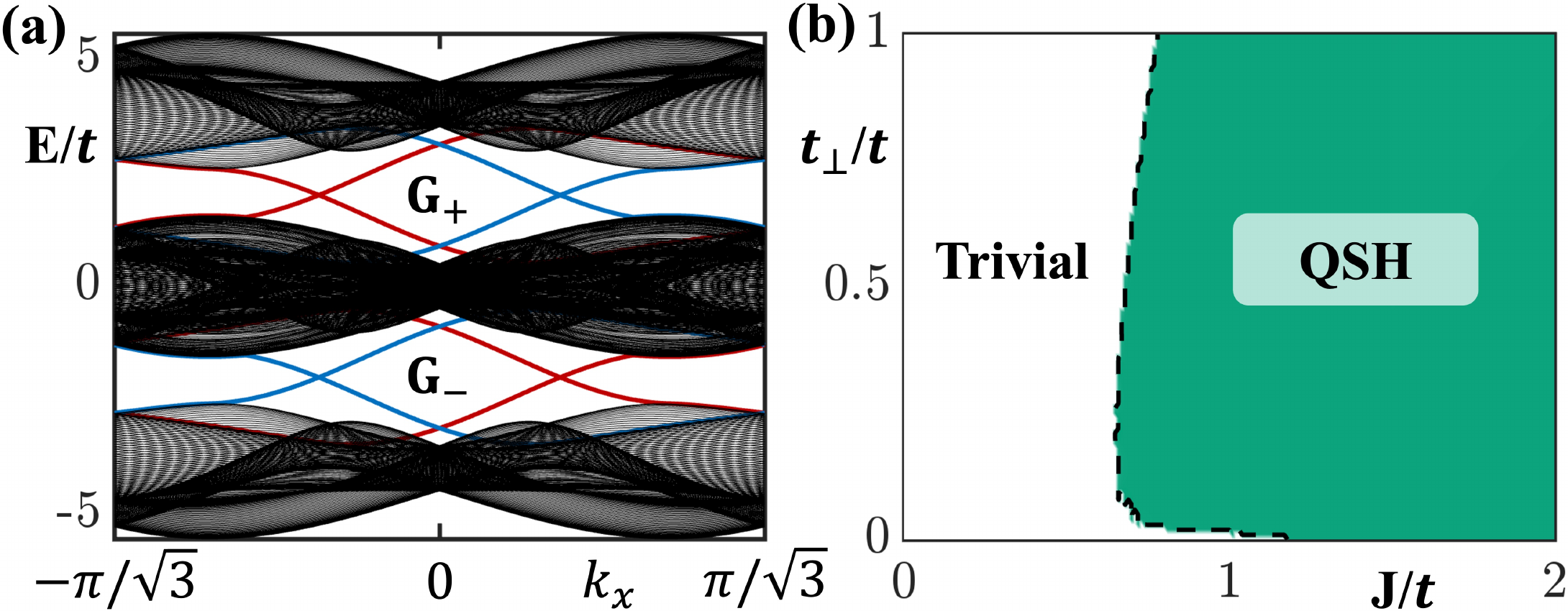}
\subsection*{Figure 3. Altermagnetic quantum spin Hall insulators}
(a) Mid-gap helical edge states at $y$-normal zigzag edges for $\{t_{0},t_{\perp},J\}=\{0.1,0.1,2\}$, providing an explicit finite-$t_0$ insulating example. (b) Direct-gap topological phase diagram as a function of interlayer coupling $t_{\perp}$ and exchange field $J$ at $t_{0}=0$. In the green region, the gaps $G_{\pm}$ carry spin Chern numbers $\mathcal{C}^{S}_{G_{\pm}}=\pm1$. This region corresponds to a QSH insulator wherever the global indirect gap is positive. Common parameters: $\{t,t_{d},\delta\}=\{1,0.5,0\}$.

\noindent\includegraphics[width=0.85\linewidth]{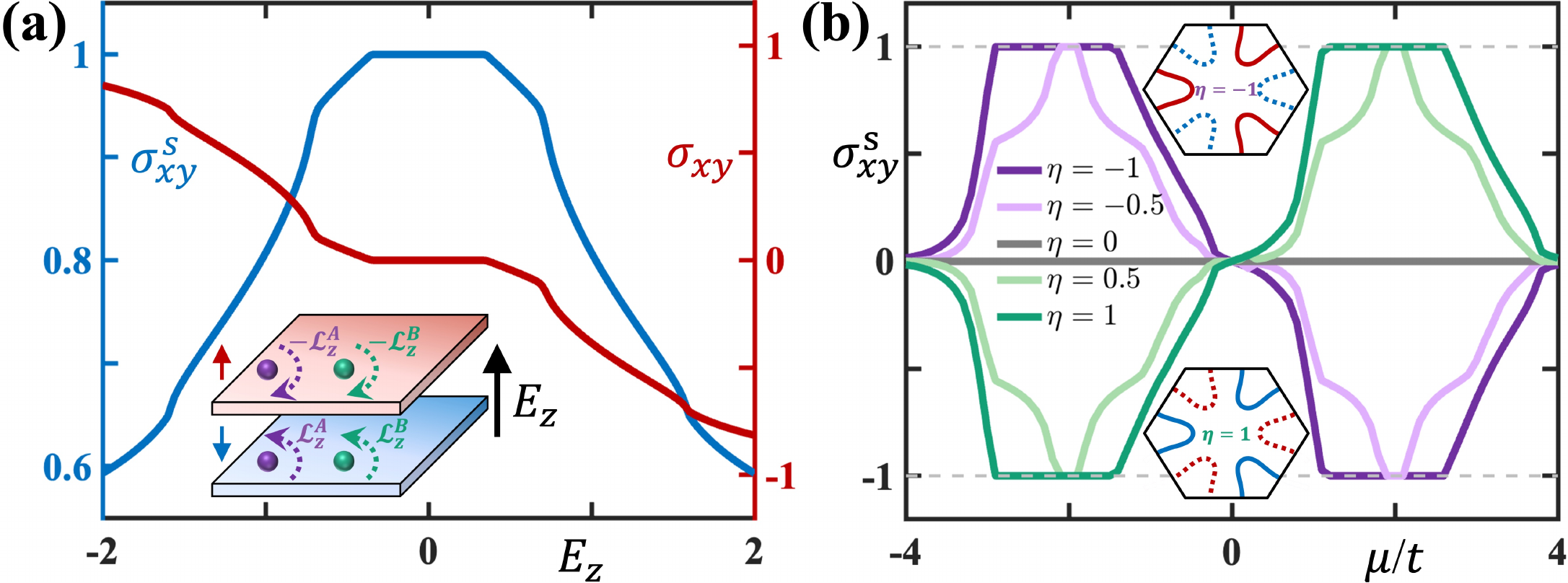}
\subsection*{Figure 4. Electric-field-induced charge Hall response and zero-temperature spin Hall conductance}
(a) Charge Hall conductance $\sigma_{xy}$ (in units of $e^{2}/h$) and spin Hall conductance $\sigma_{xy}^{s}$ (in units of $e/2\pi$) at chemical potential $\mu=1.2$. Here $E_z$ is expressed in units of $t/(ea)$, and we set $d=0.5a$.
(b) Spin Hall conductance $\sigma_{xy}^{s}$ as a function of chemical potential $\mu$ for different degrees of orbital mixing. A sign reversal occurs when $\eta$ crosses $0$. Insets: spin-resolved Fermi surfaces (red: spin-up, blue: spin-down) for $\eta=-1$ (upper) and $\eta=1$ (lower) at the same energy $E=3.25t$. Common parameters: $\{t,t_{0},t_{d},t_{\perp},\delta,J\}=\{1,0.1,0.5,0.1,0,2\}$.

\section*{MAIN TABLES, INCLUDING TITLES AND LEGENDS}

\subsection*{Table 1. Bond vectors and associated phase angles}

\noindent\resizebox{\columnwidth}{!}{%
\begin{tabular}{|c|c|c|}
	\hline
	& Bond vectors &  Complex phases $\varphi_{i}$\\
	\hline
	\multirow{4}{*}{Square} 
	& \multirow{2}{*}{\makecell{$\ba_{1}=a\{1,0\}$, $\ba_{2}=a\{0,1\}$,\\ $\ba_{3}=a\{-1,0\}$, $\ba_{4}=a\{0,-1\}$}}
	& \multirow{2}{*}{\makecell{$(i-1)\pi/2$}}\\
	&&\\
	\cline{2-3}
	& \multirow{2}{*}{$\bb_{1}=\ba_{1}+\ba_{2}$, $\bb_{2}=\ba_{2}+\ba_{3}$}
	& \multirow{2}{*}{0}\\
	&&\\
	\hline
	\multirow{5}{*}{Hexagonal} 
	&\multirow{3}{*}{\makecell{$\ba_{1}=a\{0,1\}$, $\ba_{2}=a\{\frac{\sqrt{3}}{2},-\frac{1}{2}\}$,\\ $\ba_{3}=a\{-\frac{\sqrt{3}}{2},-\frac{1}{2}\}$}}
	&\multirow{3}{*}{$(i-1)2\pi/3$} \\
	&&\\
	&&\\
	\cline{2-3}
	&\multirow{2}{*}{\makecell{$\bb_{1}=\ba_{2}-\ba_{3}$, $\bb_{2}=\ba_{3}-\ba_{1}$,\\ $\bb_{3}=\ba_{1}-\ba_{2}$}}
	&\multirow{2}{*}{0} \\
	&&\\
	\hline
\end{tabular}
}

Bond vectors and associated phase angles for inter-sublattice ($\ba_{i}$) and intra-sublattice ($\bb_{i}$) hoppings on the square and hexagonal lattices shown in Figure 2. The lattice constant $a$ is set to unity throughout.

\subsection*{Table 2. The functions $h_{i,\bk}^{p/f}$ for $p$- and $f$-wave altermagnets}

\begin{tabular}{|c|c|c|}
	\hline
	& $p$-wave &  $f$-wave\\
	\hline
	$h_{1,\bk}^{p/f}$ & $2\cos k_{x}\cos k_{y}$ & $\cos\sqrt{3}k_{x}+2\cos\frac{\sqrt{3}k_{x}}{2}\cos\frac{3k_{y}}{2}$\\
	\hline
	$h_{2,\bk}^{p/f}$ & $(t_{1}-t_{2})\cos k_{x}$ & $t(\cos k_{y}-\cos\frac{\sqrt{3}k_{x}}{2}\cos\frac{k_{y}}{2})$\\
	\hline
	$h_{3,\bk}^{p/f}$ & $(t_{1}+t_{2})\sin k_{x}$ & $t(\sin k_{y}+\cos\frac{\sqrt{3}k_{x}}{2}\sin\frac{k_{y}}{2})$\\
	\hline
	$h_{4,\bk}^{p/f}$ & $0$ & $\sqrt{3}t\sin\frac{\sqrt{3}k_{x}}{2}\sin\frac{k_{y}}{2}$\\
	\hline
	$h_{5,\bk}^{p/f}$ & $-2t_{y}\sin k_{y}$ & $-\sqrt{3}t\sin\frac{\sqrt{3}k_{x}}{2}\cos\frac{k_{y}}{2}$\\
	\hline
\end{tabular}


\bibliography{dirac}

\bigskip


\clearpage
\appendSupplementPDF{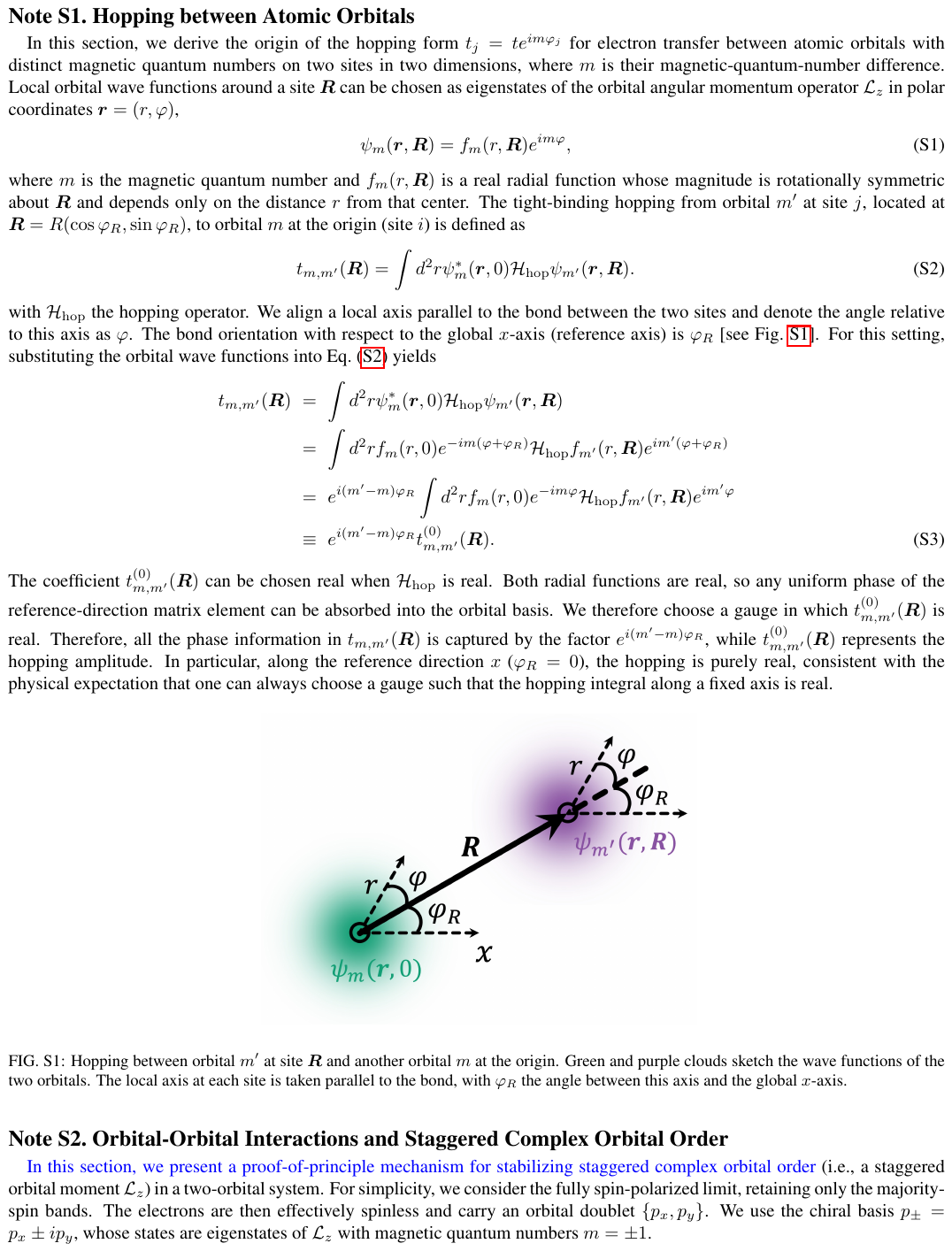}

\end{document}